\documentclass[a4paper,fleqn]{cas-dc}
\usepackage[numbers,sort&compress]{natbib}
\usepackage{multirow}
\usepackage[utf8]{inputenc}
\usepackage{graphicx}
\usepackage{hyperref}
\usepackage{epstopdf}
\usepackage{adjustbox}
\usepackage{float}
\usepackage{cancel}
\usepackage{soul}
\usepackage[normalem]{ulem} 
\usepackage{amsmath}
\def\tsc#1{\csdef{#1}{\textsc{\lowercase{#1}}\xspace}}
\tsc{WGM}
\tsc{QE}
\tsc{EP}
\tsc{PMS}
\tsc{BEC}
\tsc{DE}

\DeclareUnicodeCharacter{2212}{-}
\begin{document}
\let\WriteBookmarks\relax
\def\floatpagepagefraction{1}
\def\textpagefraction{.001}
\shorttitle{Role of shape anisotropy on thermal gradient-driven domain wall dynamics in magnetic nanowires}
\shortauthors{M. T. Islam et~al.}

\title [mode = title]{Role of shape anisotropy on thermal gradient-driven domain wall dynamics in magnetic nanowires}
\author[1]{M. T. Islam}[orcid=0000-0002-3846-4009]
\cormark[1]
\ead{torikul@phy.ku.ac.bd}
\author[1]{M. A. S. Akanda}[orcid=0000-0002-6742-2158]
\author[1]{F. Yesmin}[orcid=0000-0002-7500-1572]
\author[2]{M. A. J. Pikul}[orcid=0000-0002-9933-9922]
\author[1]{J. M. T.  Islam}[orcid=0000-0003-1723-6012]

\address[1]{Physics Discipline, Khulna University, Khulna 9208, Bangladesh}
\address[2]{Department of Physics, Colorado State University, Fort Collins, Colorado 80523, USA}
\cortext[cor1]{Corresponding author:}

\begin{abstract}
We investigate the magnetic domain wall (DW) dynamics in uniaxial$\slash$biaxial nanowires under a thermal gradient (TG). The findings reveal that the DW propagates toward the hotter region in both nanowires. The main physics of such observations is the magnonic angular momentum transfer to the DW. The hard (shape) anisotropy exists in biaxial nanowire, which contributes an additional torque, hence DW speed is larger than that in uniaxial nanowire. With lower damping, the DW velocity is smaller and DW velocity increases with damping which is opposite to usual expectation. To explain this, it is predicted that there is a probability to form the standing spin-waves (which do not carry net  energy/momentum) together with travelling spin-waves if the propagation length of thermally-generated spin-waves is larger than the nanowire length. For larger-damping, DW decreases with damping since the magnon propagation length decreases. Therefore, the above findings might be useful in realizing the spintronic (racetrack memory) devices.
\end{abstract}


\maketitle

\section{Introduction}
Efficient manipulation of domain wall (DW) in magnetic nanostructures has drawn much attention because of its potential applications in spintronic devices such as in data storage devices \cite{parkin2008, tsoi2003magnetic} and logic operations \cite{allwood2002submicrometer, allwood2005}. Several controlling parameters, such as magnetic fields, microwaves, and spin-polarized currents, have been reported in the recent decade to drive DW in magnetic nanostructures. But these parameters suffer from certain limitations in applications. Particularly, under external magnetic field, a static DW cannot exist in a homogeneous magnetic nanowire and hence a propagation of DW is obtained due to the dissipation of energy for Gilbert damping. The DW velocity is proportional to the rate of energy dissipation\cite{wang2009magnetic, wang2009high}. However, the magnetic field is unable to drive a series of DWs synchronously \cite{atkinson2003magnetic, beach2005dynamics, hayashi2006influence}. On the other hand, an spin polarized current can drive a DW, or a series of DWs in the same direction with the mechanism of spin transfer torque \cite{berger1996emission, slonczewski1996current, zhang2004roles, tatara2004theory}. Besides, a high critical current density is required to obtain a useful DW speed, which causes a Joule heating problem. \cite{hayashi2006influence, yamaguchi2004real, yamaguchi2011temperature}. To overcome such challenges, spin-wave-spin-current generated by thermal gradient (TG) emerges as an alternative driving force for the DW motion \cite{bauer2012spin, jiang2013direct}. Therefore, the study of TG-driven DW motion is significant not only for spintronic device applications but also for the understanding of the interplay between spin-wave and magnetic DW \cite{wang2012magnonic, safranski2017spin}.

Presently, to explain the physical picture of the TG driven DW dynamics, two theories of different origins are available, namely, microscopic (magnonic theory) theory \cite{yan2011all, kovalev2012thermomagnonic, wang2012domain, yan2015thermodynamic} and macroscopic (thermodynamic theory) theory \cite{schlickeiser2014role, etesami2014longitudinal,  wang2014t}. The microscopic theory predicts that more magnons are generated in the hotter part and diffuse from the hotter to the colder part, and thus a magnonic spin current is produced. While magnons pass through the DW, magnons exert a magnonic torque on the DW by transfering spin angular momentum to the DW. Thus, magnons drive the DW toward the hotter region of the nanowire, opposite to the direction of the magnonic current \cite{wang2012magnonic, kovalev2012thermomagnonic, hinzke2011domain}. The thermodynamic theory suggests that, in order to minimize the system free energy, TG produces an entropic force which drives the DW towards the hotter region \cite{schlickeiser2014role, etesami2014longitudinal}.

In our recent study \cite{islam2019t}, we demonstrated the TG induced DW dynamics in a uniaxial nanowire in which the physical mechanism is the magnonic spin transfer torque to the DW rather than energy dissipation. However,  the nanowire structure has a significant effect on DW dynamics under TG. Therefore, in this paper, we study DW dynamics along a biaxial (rectangular cross-section) and uniaxial (square cross-section) nanowire under TG and present a comparison of the DW dynamics in the considered nanowires. We find that the DW propagates toward the hotter region in both nanowires. For the uniaxial nanowire, the DW propagates towards a hotter region, accompanying a rotational motion of the DW-plane around the easy axis. Both (propagation and rotational) speeds increase proportionately with the increase of TG. For the biaxial nanowire, DW also propagates in the hotter region with a significantly larger speed than the speed in the uniaxial nanowire. DW speed in biaxial nanowire initially increases with the TG. However, after certain TGs, the propagation speed decreases and afterwards, it shows an increasing trend again for the further increment of a TG, i.e., the so-called Walker breakdown phenomenon is observed. We also study the Gilbert damping ($\alpha$) dependent DW dynamics. With lower damping, the DW velocity is smaller and DW velocity increases with damping which is opposite to usual desired. To explain this, it is predicted that there is a probability to form the standing spin-waves (which does not carry energy) together with travelling spin-waves if the propagation length of thermally-generated spin-waves is larger than the nanowire length. For larger-damping, linear and rotational speeds of DW decrease with damping since $\alpha$ reduces the magnonic current. So, the above findings of DW dynamics in biaxial$\slash$uniaxial nanowire under TG may lead to realizing the spintronics (specially racetrack memory) devices.

\begin{figure}
    \centering
	\includegraphics[width=80mm]{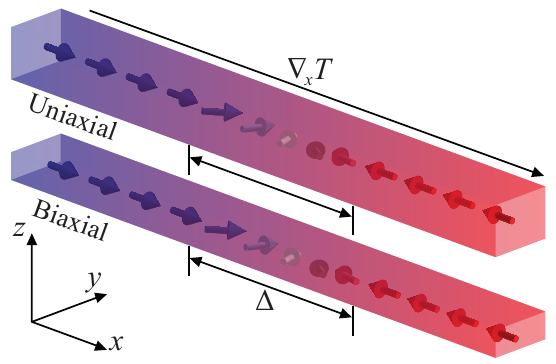}  
    \caption{Schematic diagram of considered nanowires (Uniaxial and biaxial) with a head-to-head DW at the centre under the TG $\nabla_x T$. The Blue (red) colour represents the colder (hotter) region of the sample.}
    \label{Fig1}
\end{figure}

\section{Analytical Model and Method}
We consider the nanowires of two different cross-sectional geometries, namely  square and rectangular cross-sectionals in which the DW is placed at the center of the nanowires as shown in Fig.\ref{Fig1}. $L_x$ and $L_y \times L_z$ are length and cross-sectional area of the nanowires, respectively. The DW width $\Delta$ is larger than the dimension of $L_y$, $L_z$ but much smaller than $L_x$. The nanowire with rectangular cross-section geometry possess the hard-axis anisotropy along $z$-direction which is given by  $K_{z}=1/2 \mu_oM^2_s(\mathcal{N}_z-\mathcal{N}_y)$ with $\mathcal{N}_z \ge \mathcal{N}_y$, where $\mathcal{N}_x,\mathcal{N}_y$ and $\mathcal{N}_z$ are demagnetization factors. The nanowire with rectangular cross-section is  referred as a biaxial nanowire. However, the  nanowire with square cross-section geometry possess $K_{z} = 0$, since $\mathcal{N}_z = \mathcal{N}_y$. The  nanowire with square cross-section geometry is referred as uniaxial nanowire. A TG as a driving force is applied along the nanowire. The applied temperature interval is, 100-714 K and we measure the DW dynamics for the region of temperature interval 100-550 K which is  below the Curie temperature $T_{c}$. In the atomistic level, the material parameters, for example, the exchange constant (which originates from the Pauli exclusion principle) and the crystalline anisotropy (which originates from the spin–orbit coupling) weakly relay on the temperature because of the vibration of atoms \cite{chico2014thermally}. So, it is assumed that the saturation magnetization and exchange constant are temperature-independent \cite{islam2019t}. The magnetization dynamics is governed by the stochastic Landau-Lifshitz-Gilbert (sLLG) equation \cite{brown1963thermal, gilbert2004phenomenological, vansteenkiste2014design},
\begin{equation}
    \dfrac{d\textbf{m}}{dt} = -\gamma \textbf{m}\times(\textbf{h}_{\text{eff}}+ \textbf{h}_{\text{th}})+\alpha\textbf{m}\times\dfrac{\partial\textbf{m}}{\partial{t}}
    \label{sllg}
\end{equation}
where $\mathbf{m}=\mathbf{M}/M_{s}$ and $M_{s}$ are respectively the magnetization direction and the saturation magnetization, $\alpha$ is the Gilbert damping constant, $\gamma$ is the gyromagnetic ratio. $\mathbf{h}_\text{eff} = \frac{2A}{\mu_0 M_s} \sum_\sigma \frac{\partial^2 \mathbf{m}}{\partial x_\sigma^2} + \frac{2 K_x}{\mu_0 M_s} m_x \hat{\mathbf{x}} + \mathbf{h}_\text{dipole}$ is the effective field, where $A$ is the exchange stiffness constant, $x_\sigma$ ($\sigma=1,2,3$) denote Cartesian coordinates $x$, $y$, $z$, $K_x$ is the easy-axis anisotropy, $\mathbf{h}_\text{dipole}$ is the dipolar field and $\mathbf{h}_\text{th}$ is the thermal stochastic field.

The stochastic LLG equation is solved numerically by the micromagnetic simulation package MUMAX3 \cite{vansteenkiste2014design} in which we have used adaptive Heun solver. The time step is chosen $10^{-14}$s for the unit cell size ($2\times2\times2$) $\text{nm}^{3}$ and $10^{-15}$s for unit cell smaller than ($2\times2\times2$) $\text{nm}^{3}$. The saturation magnetization $M_\text{s} = 8 \times10^{5} \text{A}/\text{m}$ and exchange stiffness constant $A = 13\times10^{-12} \text{J}/\text{m}$ are used to mimic permalloy in our simulations. The thermal field follows the Gaussian process characterized by following statistics \cite{islam2019t, hinzke2008domain, wang2016thermally} 

\begin{equation}
    \begin{gathered}
        \langle h_{\text{th},ip}(t) \rangle = 0, \\
        \langle h_{\text{th},ip}(t)h_{\text{th},jq}(t+\Delta t) \rangle = \frac{2k_\text{B}T_i \alpha_i}{\gamma M_\text{s} a^3} \delta_{ij} \delta_{pq} \delta(\Delta t)
    \end{gathered}
\end{equation}
where $i$ and $j$ denote the micromagnetic cells, and $p$, $q$ represent the Cartesian components of the thermal field. $T_{i}$ and $\alpha_i$ are respectively temperature and the Gilbert damping at cell $i$, and $a$ is the cell size. $k_\text{B}$ is the Boltzmann constant. The numerical results presented in this study are averaged over $15$ random configurations.
With the time step $\Delta t$, the thermal/random field, which is related to temperature, can be expressed as
\begin{equation}
    h_{\text{th, i,p}} = \eta \sqrt{\dfrac{2\alpha k_\text{B}T}{\gamma M_\text{s}a^3\Delta t}}
\end{equation}
where $\eta$ is a therm seed (random number) which gives the normal distribution with zero average.

\section{Numerical Results}

\begin{figure}
    \centering
	\includegraphics[width=80mm]{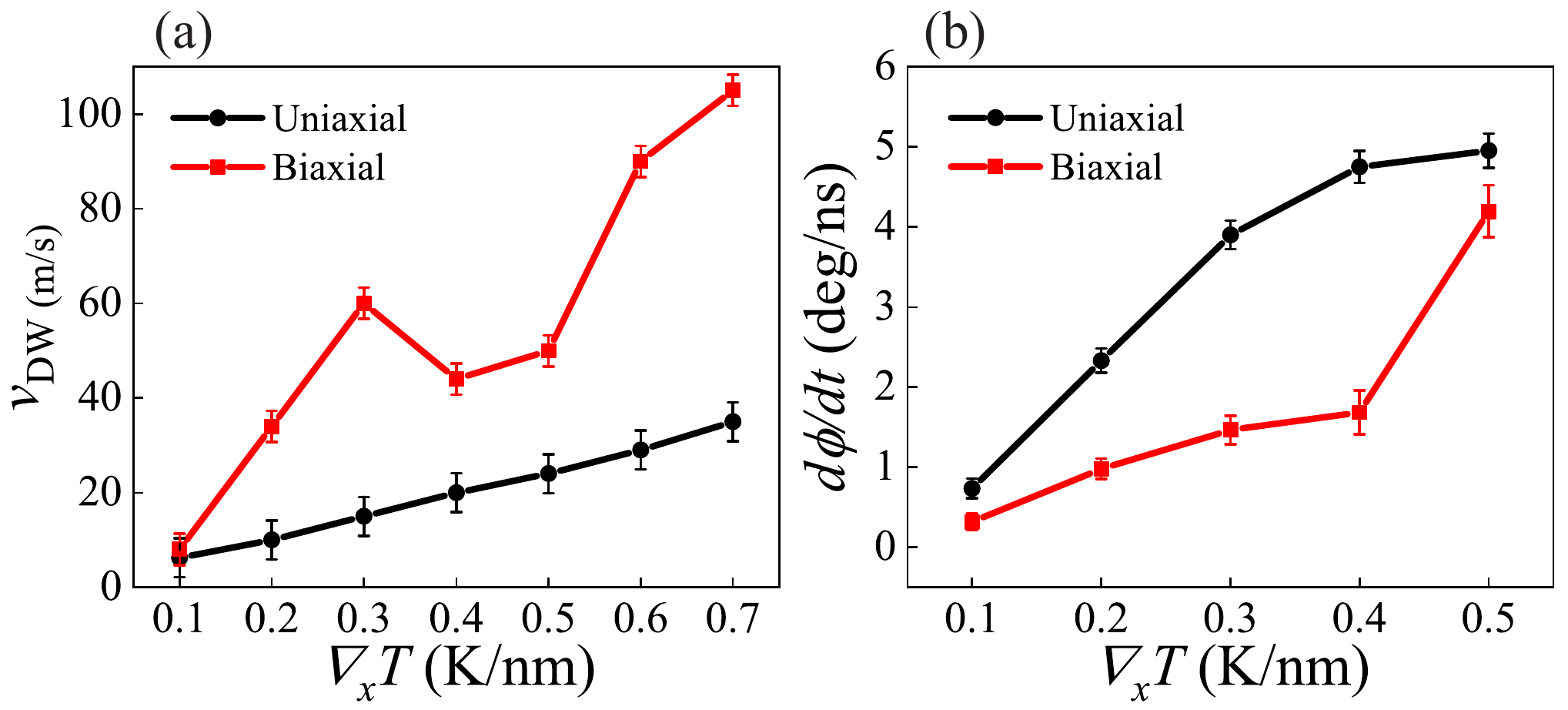}  
    \caption{A comparison of DW dynamics under TG  $\nabla_{x}T$ along biaxial ($1024\times30\times4$ $\text{nm}^{3}$) and uniaxial ($ 1024\times4\times4$ $\text{nm}^{3}$) nanowire. (a)  Red and black lines represent the DW-linear speeds ($v_\text{DW}$) along biaxial and uniaxial nanowire, respectively. (b) Black and red lines represent rotational speeds of DW-plane ($\dfrac{d\phi}{dt}$) along uniaxial and biaxial nanowire, respectively.}
    \label{Fig2}
\end{figure}

\subsection{Thermal gradient dependent DW dynamics}

In order to compare the DW dynamics (linear and DW-angular speeds) in the biaxial and uniaxial nanowires, $15$ random trajectories of DW propagation are generated for each set of parameters and the TG. Then we take the statistical ensemble average to obtain the time dependent average DW position, and hence linear ($v_\text{DW}$) and rotational (${d\phi}/{dt}$) speeds of DW are estimated. In both cases, the DWs propagate toward the hotter region. The comparison between linear and rotational speeds of DW in biaxial and uniaxial nanowires is shown in Fig. \ref{Fig2}(a) and (b), respectively. In the uniaxial nanowire under TG, the DW moves toward the hotter regions accompanying the rotation of the DW plane around the symmetry$\slash$easy axis. DW speed proportionately increases with the TG. It is interesting to note that the DW linear speed (rotational speed) in the biaxial nanowire is significantly larger (smaller) than the DW linear speed (rotational speed) in the uniaxial nanowire. To explain the reason why the DW linear speed is larger in biaxial nanowire, one can look at the expression of hard anisotropy $K_z = 1/2 \mu_0 M^2_s(\mathcal{N}_z-\mathcal{N}_y)$, say along $z$-direction for biaxial nanowire while $K_z=0$ for uniaxial nanowire (since $\mathcal{N}_z$ and $\mathcal{N}_y$ are equal). This hard-axis anisotropy contributes a polar torque ($\Gamma_{\theta}$) along $\theta$-direction \cite{mougin2007d} since $\Gamma_{\theta}$ is proportional to $(\mathcal{N}_z-\mathcal{N}_y)$.

After certain TG, the DW-linear speed (rotational speed) decreases (increases), which indicates the Walker breakdown phenomena. With the further increment of TG, both DW-linear and rotational speeds show increasing trends. After the Walker breakdown limit, the biaxial nanowire behaves like a uniaxial nanowire and the DW linear and rotational speeds increase similarly to the uniaxial nanowire. These DW dynamics can be explained by microscopic theory as stating that more magnons are populated in the hotter region and travel from the hotter to the colder region. During crossing the DW, the magnon exerts a magnonic torque on the DW by transferring spin angular momentum to the DW. In this way, magnons drive the DW toward the hotter region of the nanowire, opposite to the direction of the magnonic current \cite{yan2011all, kovalev2012thermomagnonic, wang2012domain, yan2015thermodynamic}. These findings are also consistent with the macroscopic theory \cite{schlickeiser2014role, etesami2014longitudinal}. 

\begin{figure}
    \centering
	\includegraphics[width=80mm]{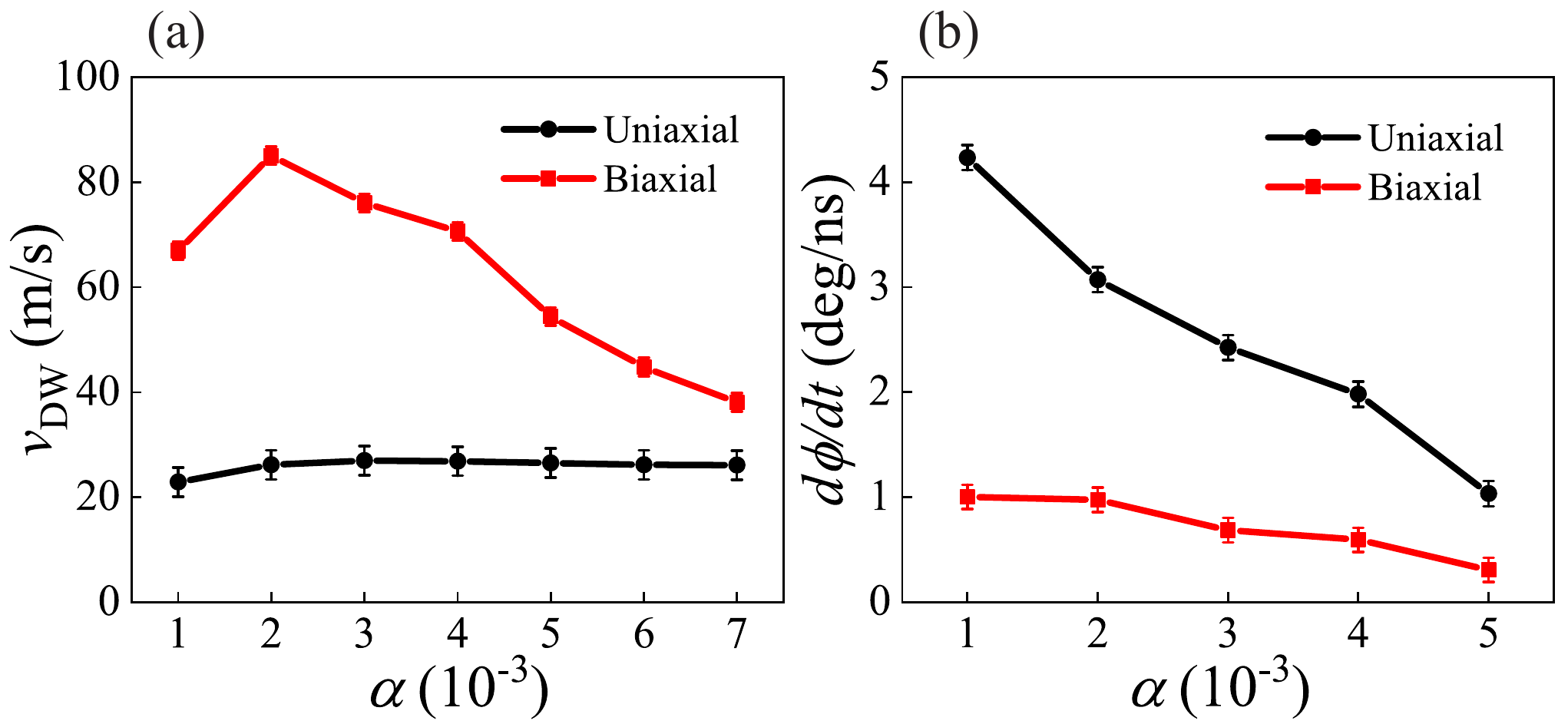}  
    \caption{Damping ($\alpha$) dependent (a) DW-linear speed represented by a red line (black line) along biaxial (uniaxial) nanowire. (b) Rotational speed of DW-plane is represented by a red line (black line) along biaxial (uniaxial) nanowire.}
    \label{Fig3}
\end{figure}

\subsection{Damping dependent DW dynamics}
Gilbert damping coefficient $\alpha$ is a crucial material parameter that has a significant effect on DW dynamics under TG. For fixed  $\nabla_{x}T =0.5$ $\text{K}/\text{nm}$, we investigate the DW dynamics in biaxial and uniaxial nanowires of length $L_x= 1024$ nm. The obtained results of linear and rotational speeds of DW as a function of $\alpha$ are presented as in Fig. \ref{Fig3}(a) and (b), respectively. In both nanowires, DW speed increases initially and then decrease with increasing $\alpha$. Usually, the DW velocities should be decreased since the magnon propagation length decreases with damping. However, we observe that the DW speeds increase initially till a certain $\alpha$ and then decrease monotonically with $\alpha$. To explain this observation, one can recall the basic physics of generating standing waves in the nanowire with fixed boundaries. For lower damping, propagation of thermally generated spin waves is larger and if it is larger than the nanowire length. It is predicted that there is a probability to form the standing spin waves together with travelling spin waves in the nanowire. On the other hand,  the standing spin-wave does not carry any energy$\slash$spin angular momentum while only travelling spin wave does. Consequently, the energy transfer to the DW is decreased, i.e., DW velocity is decreased due to the decrement of the magnonic current. However, for larger $\alpha$, the formation of standing spin-waves decreases due to the reduction of propagation length of spin waves and oppositely the number of travelling spin wave increases. Therefore, the spin-wave-spin-current effectively, constitutes from travelling spin waves,  increases, and hence the DW speeds increases. The extensive explanation requires to distinguish the stationary and travelling spin wave modes from simulation data and hence to calculate the net spin-wave-spin current density for obtaining DW velocity  which might be a scope of future studies. For the large $\alpha \ge 0.002$, with the increment of $\alpha$, the magnon propagation length decreases with the damping, so the amount of spin angular momentum deposited on a DW should decrease. As a result, the DW propagation speed and DW-plane rotation speed decrease with $\alpha$ as shown in Fig. \ref{Fig3}(a) and (b), respectively. For fixed $\alpha$, TG and other material parameters,  it is predicted that the DW speeds will be decreased monotonically with  $\alpha$ if the length of the nanowire is chosen large enough to rule out the standing spin wave. The above statement would be justified in the next subsection.

\begin{figure}
    \centering
	\includegraphics[width=80mm]{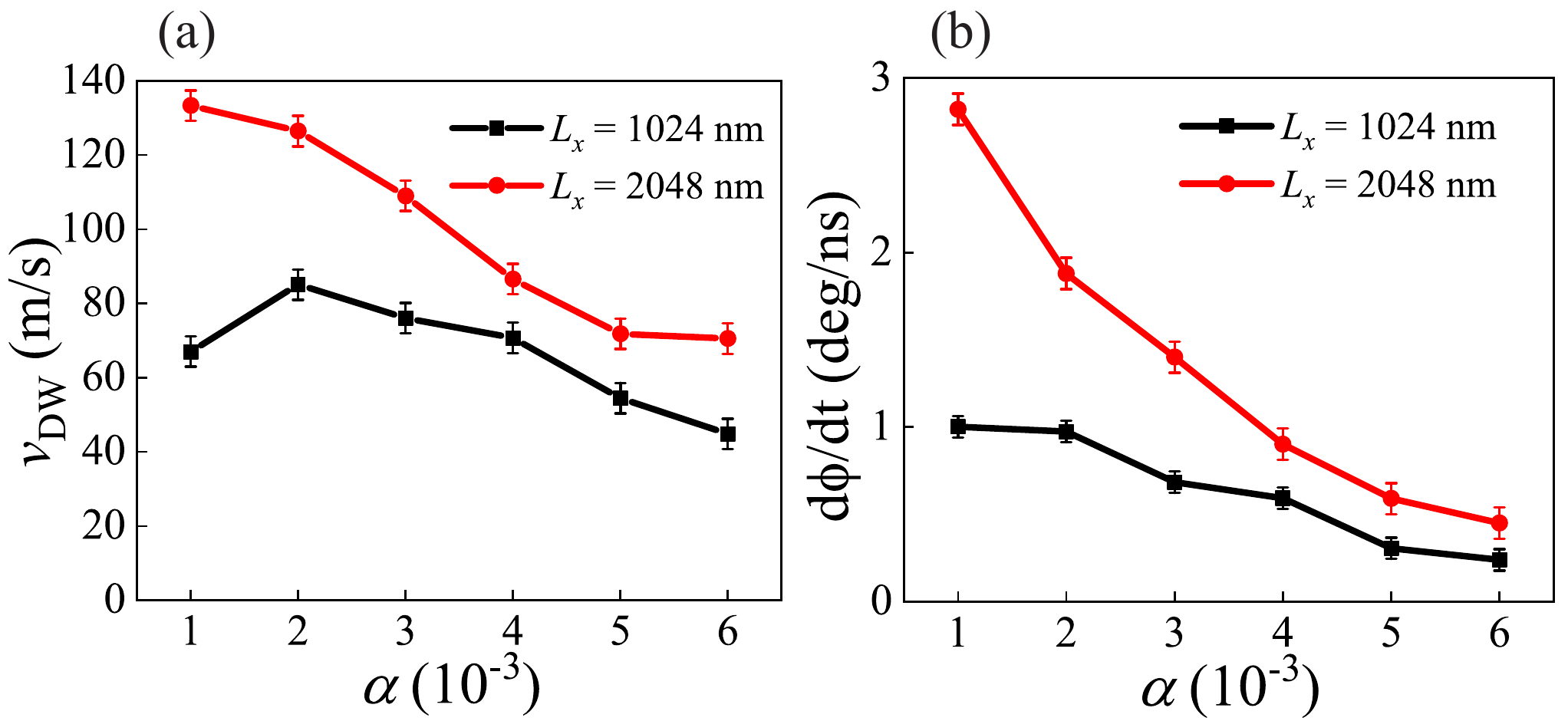}  
    \caption{Damping $\alpha$ dependent (a) DW-linear speed represented by black line (red line) along biaxial nanowire of length $L_x=1024$ nm ($L_x=2048$ nm). (b) Rotational speed of DW-plane represented by black line (red line) along biaxial nanowire of length $L_x=1024$ nm ($L_x=2048$ nm).}
    \label{Fig4}
\end{figure}

\subsection{Boundary effect on DW dynamics}
Here, we present the behaviour of DW dynamics in two different lengths i.e., $L_{x}=1024$ nm and $L_{x}=2048$ nm of biaxial nanowire in Fig. \ref{Fig4}(a) and (b), respectively, as a function of $\alpha$ (ranging from 0.001 to 0.006) for fixed TG $\nabla_{x}T = 0.5$ $\text{K}/\text{nm}$. Fig. 4(a) shows the DW linear velocity as a function of $\alpha$ corresponding to the nanowire lengths, $L_{x}=1024$ nm (black) and $L_{x}=2048$ nm (red). In Fig. \ref{Fig4}(b), the black line and red line show the change of angular velocity as a function of $\alpha$ corresponding to the lengths $L_{x}=1024$ nm and $L_{x}=2048$ nm, respectively. DW angular velocities also decrease with $\alpha$. Since the magnons propagation length decreases with $\alpha$ \cite{hinzke2011domain}, it reduces the number of magnons to pass through the DW, which leads to the reduction of DW velocity proportionately. It is noted that for smaller $L_{x}=1024$ nm (Fig. \ref{Fig4}(a)), the DW speed is lower for $\alpha = 0.001$, and it gradually increases and then monotonically decreases. Interestingly, such behaviour disappears with  for the larger length $L_{x}=2048$ nm and the DW velocities show decreasing trend monotonically. The reason for observing such behaviour has been discussed above for shorter $L_{x}=1024$ nm. Here we mention again shortly that for lower damping, the magnon propagation length is larger. If the length of the sample size is not large enough compare to the distance of two ends of nanowire, then there is a probability of forming the standing waves of some modes between the fixed boundaries. Consequently, the energy transfer through the DW is decreased, i.e., DW velocity is decreased because of the reduction of the magnonic current.

\begin{figure}
    \centering
	\includegraphics[width=80mm]{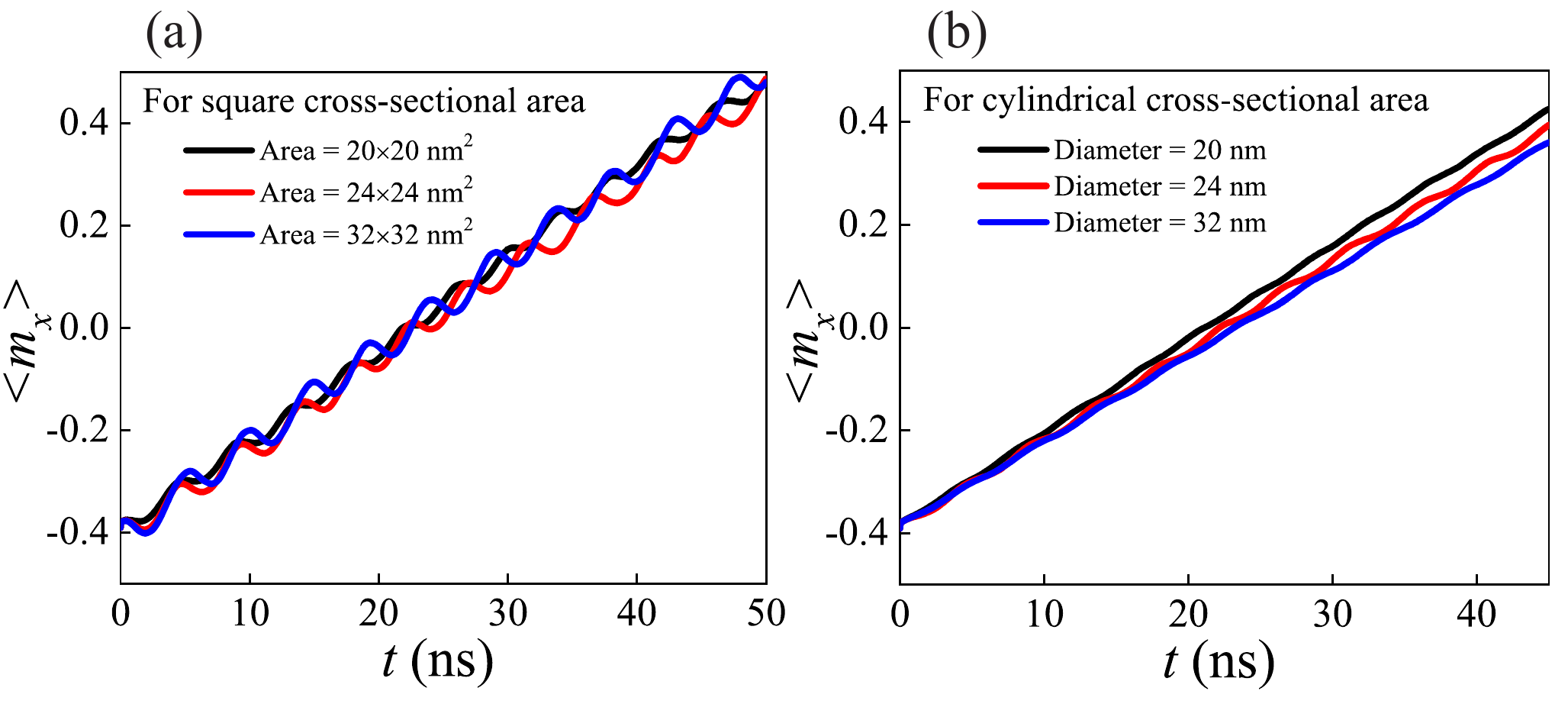}  
    \caption{(a) Square and (b) cylindrical cross-sectional area  dependent DW propagation along uniaxial nanowire.}
    \label{Fig5}
\end{figure}

\subsection{Cross-sectional area dependent DW dynamics}
The cross-sectional area of the nanowire has an influence on the DW dynamics. Fig. \ref{Fig5}(a) and (b) show the DW dynamics in uniaxial nanowires of square and cylindrical cross-sectional area, respectively. For the nanowire with square cross-sectional area, it is observed that the DW propagates towards the hotter region with oscillatory or wavy nature. For square cross-sectional area (larger) ($32\times32$) nm$^2$, the oscillation amplitude is larger. This is because of the energy changes due to deformations of the DW shape while DW rotates around the easy axis of the nanowire \cite{altbir2020tuning}. So the DW propagates with wavy nature. This wavy nature of DW propagation decreases in the sample of cylindrical cross-section as shown in Fig. \ref{Fig5}(b). The reason is that, in the sample of cylindrical cross-section, the DW experiences almost symmetric shape anisotropy. Fig. \ref{Fig5}(b) shows still small wavy nature propagation in the sample of cylindrical cross-sectional area because it is technically difficult to simulate the perfect cylindrical cross-section.

\section{Discussions and Conclusions}
In this study, we investigate the DW dynamics in uniaxial and biaxial nanowires under TG. We found that the DW propagates toward the hotter region in both nanowires. For the uniaxial nanowire, the DW propagates towards a hotter region, accompanying a rotational motion of the DW-plane around the easy axis. Both (propagation and rotational) speeds increase proportionately with the increase of TG. For biaxial nanowire, DW also propagates in the hotter region, and DW propagation speed is significantly larger than the speed in the uniaxial nanowire. DW speed in biaxial nanowire initially increases with the TG. But after certain TGs, the propagation speed decreases and afterwards, it shows an increasing trend again for the further increment of a TG, i.e., the so-called Walker breakdown phenomenon is observed. We also study the Gilbert damping $\alpha$ dependent DW dynamics. With lower damping, the DW velocity is smaller and DW velocity increases with damping which is opposite to usual desired. To explain this, it is predicted that there is a probability to form the standing spin-waves (which does not carry energy) together with travelling spin-waves if the propagation length of thermally-generated spin-waves is larger than the nanowire length. For larger damping, linear and rotational speeds of DW decrease with damping since $\alpha$ reduces magnon propagation length and thus reduces the magnonic current. In this study, we also examined the length and cross-sectional area dependent DW dynamics with damping coefficient. We found that it is required to optimize the damping coefficient with the nanowire structure to obtain the efficient DW speed under TG. Therefore, the above findings of DW dynamics in biaxial$\slash$uniaxial nanowire under TG may might be useful for the fundamental interests and applications in spintronics devices.

\section*{Acknowledgments}
This research was supported by the Khulna University Research Cell (Grant No. KU/RC-04/2000-158), Khulna, Bangladesh and the Ministry of Education (BANBEIS, Grant No. SD2019972).

\bibliographystyle{elsarticle-num}
\bibliography{bibliography}
\end{document}